\documentclass[preprint,preprintnumbers,amsmath,amssymb]{revtex4}
\usepackage{graphicx}% Include figure files
\usepackage{dcolumn}% Align table columns on decimal point
\usepackage{bm}% bold math

\begin{document}

\title{Non-Gaussianity effect of petrophysical quantities by using q-entropy and multi fractal random walk}
\author{Z. Koohi lai $^1$, S. Vasheghani Farahani $^2$, G. R. Jafari $^3$ \\
{\small $^1$ Plasma Physics Research Center, Science and Research
Branch, Islamic Azad university, Tehran, Iran} \\
{\small $^2$ Department of Physics, Tafresh University, Tafresh, Iran} \\
{\small $^3$ Department of Physics, Shahid Beheshti University,
G.C., Evin, Tehran 19839, Iran}}

\begin{abstract}
The geological systems such as petroleum reservoirs is investigated
by the entropy introduced by Tsallis and
multiplicative hierarchical cascade model. When non-Gaussianity
appears, it is sign of uncertainty and phase transition, which could
be sign of existence of petroleum reservoirs. Two important
parameters which describe a system at any scale are determined; the
non-Gaussian degree, $q$, announced in entropy and the
intermittency, $\lambda^2$, which explains a critical behavior in
the system. There exist some petrophysical indicators in order to characterize a
reservoir, but there is vacancy to measure scaling information contain in comparison with together, yet.
In this article, we compare the non-Gaussianity in three selected indicators in various scales.
The quantities investigated in this article
includes Gamma emission (GR), sonic transient time (DT) and Neutron
porosity (NPHI). It is observed that GR has a fat tailed PDF at all
scales which is a sign of phase transition in the system which
indicates high $q$ and $\lambda^2$. This results in the availability
of valuable information about this quantity. NPHI displays a scale
dependence of PDF which converges to a Gaussian at large scales.
This is a sign of a separated and uncorrelated porosity at large
scales. For the DT series, small $\lambda^2$ and $q$ at all scales
are a hallmark of local correlations in this quantity.

Keywords: Tsallis entropy, Non-Gaussian degree, Intermittency.
%PACS: {02.50.-r, 92.60.Sz}
\end{abstract}
\maketitle

\section{Introduction}

In petroleum reservoirs in order to find the lithology, the
well-logging technique has proved adequate for oil exploration and
production. Parallel to the research of oil and gas fields, some
features of petrophysical quantities obtained by well-logging could
be analyzed and described in terms of the kind and content of the
fluids within the pores. The quantities investigated in this
research includes Gamma emission (GR), which is a measure of the
natural radiation of the formation; sonic transient time (DT), which
is a recording of the time required for a sound wave to travel
through a formation; Neutron porosity (NPHI), which uses high-energy
Neutrons that collide with various atoms of both the formation
material and fluids, reporting the existence of Hydrogen in the pore
space \cite{Fedi,log1,log2}. These help us insight into the spatial
heterogeneity of the properties of the large scale porous media,
such as porosity, density, and the lithology at distinct length
scales \cite{Jafari,log3,log4}.

In geological systems external forces besides internal instabilities
cause the system to become a complex system. In order to analyze
geological systems the theory of complex systems needs to be
implemented. In this theory the Probability density function (PDF),
entropy, and the degree of non-Gaussianity is used. However the
working parameter in this study is the entropy which is an important
bases in thermodynamics. Entropy was introduced in $1865$ by Rudolf
Julius Emmanuel Clausius \cite{a}. Later Boltzmann showed that
entropy could be expressed in terms of the probabilities related to
the microscopic structures of the system \cite{b,c}.

When a system is in contact with a large reservoir, the
Boltzmann-Gibbs entropy is obtained as:
\begin{equation}
S_{BG}=-k\sum_{i=1}^{w} P_i \ln P_i,
\end{equation}
where $P_i$ is the probability of the microscopic configuration $i$
\cite{d}, and $k$ is the Boltzmann constant. So one can conclude
that entropy is subject to the probabilities of possibilities in
system, reflecting the information upon the physical system
\cite{z}.

It is well-known in the classical Boltzmann-Gibbs (BG) statistical
mechanics that the Gaussian PDF under appropriate constraints,
maximizes the BG entropy. BG statistics has been successful in
explaining the behavior of systems in which short spatial/temporal
interactions are significant \cite{k}. Tsallis in $1988$ proposed an
entropy for systems that may have a multifractal, scale-free or
hierarchical structure in the occupancy of their phase space as in
the form \cite{f}
\begin{equation}
S_{q}=k\frac{1-\sum_{i=1} ^{w} {P_i ^{q}}} {q-1},
\end{equation}
which generalizes $S_{BG}(\lim_{q\rightarrow {1}} S_q=S_{BG})$
\cite{g}. Where $S_{q}$ is nonnegative, concave, experimentally
robust (or Lesche-stable \cite{h}) and yield a finite entropy
production per unit time \cite{i,j}. Note that the applicability of
Tsalis entropy becomes important for systems with long-ranged
correlations which involve strong interactions, glassy systems,
fractal processes, some types of dissipative dynamics and other
systems that in some way violate ergodicity.

The entropy of expression $(2)$ has been applied to a wide range of
science such as physics, biology, chemistry, economics, geophysics
\cite{Telesca1,Telesca2,Telesca3} and medicines \cite{k} and
references therein.

\begin{figure}[t]
\centering
\includegraphics[width=10cm,height=5.5cm,angle=0]{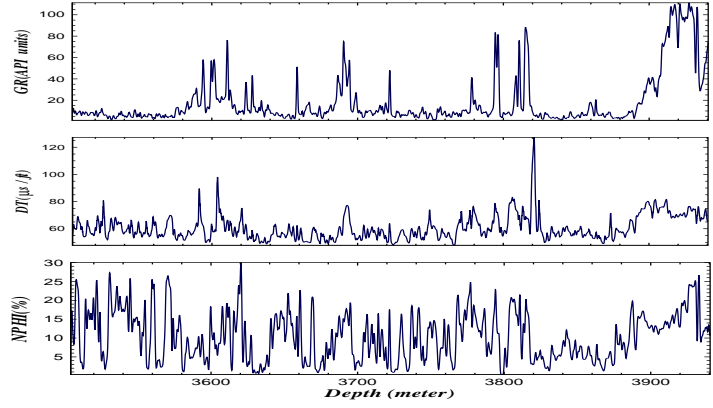}
\caption{Three petrophysical quantities from up to down: Gamma ray
(GR), sonic transient time (DT), Neutron porosity (NPHI) and depth
that measured in space interval $15.4 \, cm$ at depth $3504.5 \, m$
to $3946.8 \, m$ of a gas well in southwest of Iran.}\label{fig1}
\end{figure}
We use well log data from a gas well in southwest of Iran. These
data are Gamma ray radioactivity (GR), sonic transient time (DT) and
Neutron porosity (NPHI) taken every $15.4 \, cm$ at the depth
interval of $3504.5-3946.8$ meters. The logged interval includes
Asmari formation which is one of the main oil producer formations in
Iran reservoirs and consists mainly of fractured carbonate, sand
stone, shaly sand and a trace of anhydrate. Fig. $1$ shows GR, DT
and NPHI data sets from the depth, $d$, of $3504.5 \, m$ to $3946.8
\, m$. Our analysis is based on the increment of the petrophysical
quantities shown by $\varphi(d)$ in various scales, $s$, which
$x(s)=\varphi(d+s)-\varphi(d)$.

In this article we describe the petrophysical quantities using the
PDF and non-Gaussian degree, $q$, based on Tsallis entropy in
section $2$. We explain the connection between entropy (less of
information) and the non-Gaussian factor $\lambda^{2}$
(intermittency) for these data sets in section $3$ and the
conclusions are stated in section $4$.

\begin{figure}[t]
\centering
\includegraphics[width=15cm,height=5cm,angle=0]{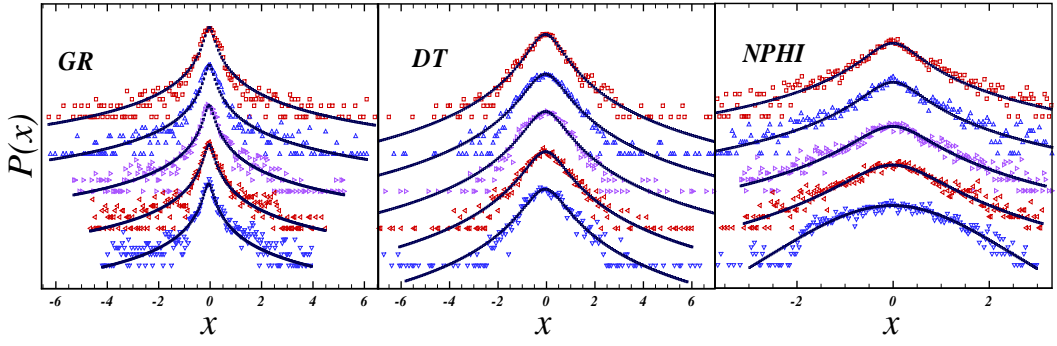}
\caption{ Increment petrophysical quantities PDF in various scales
(from top to bottom) $s=15,75,225,450,900 \, cm $. These estimations
are fulfilled on petrophysical quantities such as Gamma emission
(GR), sonic transient time (DT) and Neutron porosity (NPHI) that
measured in space interval $15.4 \, cm$ from a gas well in southwest
of Iran. Solid lines are theoretical estimation achieved by fitting
data sets to Eq. $(5)$.}\label{fig2}
\end{figure}

\section{probability distribution function using entropy $S_q$}

Since the Shannon entropy is the logarithm of the probability, the
entropy is extensive. As known the Shannon entropy is considering
all phenomenon based on the Boltzmanian probability distribution
function (PDF). While, in general the experimental systems have no
reason to obey the Boltzmanian distribution. Tsallis added a power
of $q$ to the probability to be able to generalize the Shannon
Entropy to all phenomenon, this causes the entropy not to be
extensive. As the value of $q$ increases, the effects of the tails
become more pronounced. Hence, in systems where $q$ is greater than
unity the minority is more pronounced compared a Gaussian ($q=1$)
system. In other words having the value of $q$ greater than unity,
non-Gaussianity appears.

In order to determine the non-Gaussian degree of the system, $q$, it
is essential to determine the probability distribution function
analytically. Following the formalism of \cite{k,m}, we obtain the
probability distribution function. Since the most probable
distribution corresponds to the maximum entropy, the variational
principle is applied to $S_q$. The continuous version of $S_q$ is
stated as:
\begin{equation}
S_q=k\frac {1-\int{[P(x)]^{q} dx}} {q-1},
\end{equation}
where $\int{P(x) dx}=1$ is the natural constraint corresponding to
the normalization of Eq. $(3)$ and
\begin{equation}
\int{x \frac {[P(x)]^{q}} {\int{[P(x)]^{q} dx}} dx}\equiv \langle x
\rangle_{q} =\overline{\mu}_{q}, \,\,\,\,\
\int{(x-\overline{\mu}_{q})^{2} \frac {[P(x)]^{q}} {\int{[P(x)]^{q}
dx}} dx}\equiv \langle (x-\overline{\mu}_{q})^{2} \rangle_{q}
=\overline{\sigma}_{q},
\end{equation}
are the natural constraints corresponding to the generalized and
mean variance of $x$ respectively, see \cite{k}.

\begin{figure}[t]
\centering
\includegraphics[width=13cm,height=5.5cm,angle=0]{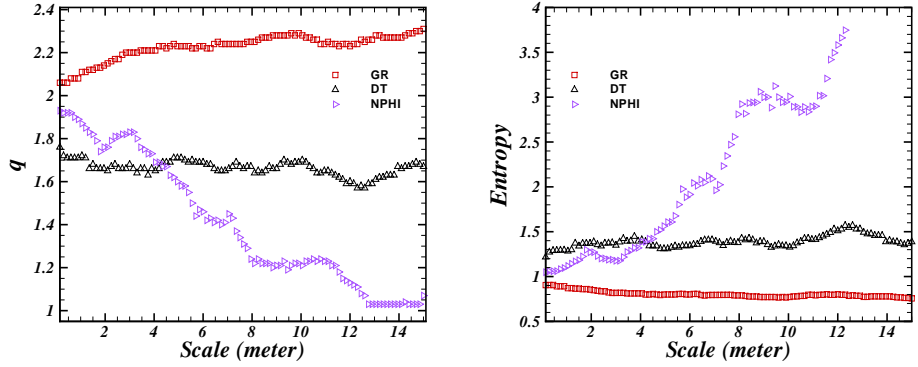}
\caption{Left panel: the scale dependence of non-Gaussian degree,
$q$, for the $3$ data sets. This estimation has been made by
determining the PDF based on Tsallis entropy then exert $\chi^2$
test method. Right panel: representation of entropy vs. the scales
by employing the definition of Tsallis entropy Eq. $(3)$ for GR
(square), DT (delta) and NPHI (right triangle).}\label{fig3}
\end{figure}
Implementing the Lagrange method in order to find the optimizing
distribution under the constraints, we obtain \cite{k}
\begin{equation}
P(x)=A_{q}[1+(q-1)B_{q}(x-\overline{\mu}_{q})^{2}]^{\frac {1} {1-q}}
\ ,\ (q<3),
\end{equation}
where
\begin{equation}
A_{q}=\left\{
        \begin{array}{ll}
          \frac {\Gamma (\frac {5-3q} {2-2q})} {\Gamma (\frac{2-q} {1-q})
}\sqrt{\frac {1-q} {\pi} B_{q}} & \ \ \hbox{$q<1$} \\
          \frac {\Gamma (\frac {1} {q-1})} {\Gamma (\frac{2-q} {2q-2})
}\sqrt{\frac {q-1} {\pi} B_{q}} & \ \ \hbox{$q>1$}
        \end{array}
      \right.
    \ \ \   \& \ \ \ B_q=[(3-q)\overline{\sigma} _{q} ^{2}]^{-1}.
\end{equation}

In order to characterize the non-Gaussian degree of the system, $q$,
the $\chi^2$ test method is employed. If $q$ is close to $1$, the
PDF tends to be Gaussian, hence, an uncorrelated series is expected.
On the other hand, the large value of the non-Gaussian factor $q$,
illustrates strong correlations and similarity of neighbors in data
sets which affects the behavior of the system. Therefore it can be
concluded that high $q$ is a hallmark of having information about
the system. For any $q$ a PDF could be plotted as shown in Fig. $2$.
In Fig. $2$ the PDF is plotted using Eq. $(5)$ (solid curves) and
the data sets (symbols), showing a satisfactory consistency. To
state more specifically, in order to estimate $q$, we use the
likelihood method which works by minimizing the parameter $\chi^{2}$
which defined as:
\begin{equation}
\chi^{2}(q)=\sum_{x}\frac{[P_{emperical}(x)-P(x,q)]^{2}}{\sigma^{2}_{emperical}(x)+\sigma^{2}_{q}(x)},
\end{equation}
where $P_{empirical}(x)$ is computed directly from the empirical
series and $P(x,q)$ is given by Eq. (5). $\sigma_{empirical}$ is the
mean standard deviation of $P_{empirical}(x)$ and $\sigma_{q}$ is
associated with the probability density function derived by Eq. (5).
The global minimum value of $\chi^{2}$ corresponds to the best value
of $q$. The best estimation of $q$, corresponding to Eq. (5), yields
the PDF which fits well to the empirical PDF.

The entropy index $q$ versus the scales is shown in Fig. $3$ (left
panel) for any specific petrophysical series. Note that the system
is described by a specific $q$ for any scale. Taking a close look at
Fig. $2$ and Fig. $3$ we deduce that the non-Gaussian PDF
corresponds to large $q$, specially at small scales. This means that
$q$ is showing the efficiency of non-Guassianity. As seen in the
three panels of Fig. $2$, non-Gaussianity of GR is more pronounced
in comparison to DT and NPHI. Hence, powerful correlations and large
fluctuations in the series are observed which influence the other
quantities and gives an insight to the formation of the system.
However, it could be seen in the third panel which is for NPHI, that
the PDF is non-Gaussian at small scales but tends to be Gaussian at
large scales which corresponds to large and small values of $q$
respectively. As the values of $q$ in NPHI tends to $1$, no
correlation exists, hence no information about the porosity would be
available \cite{Jafari}.

In the right panel of Fig. $3$, the entropy of the data sets are
plotted vs. the scales. It is shown that the entropy of GR has its
least value at all scales, meaning that we have fine information
about the existence of shaly layers detected by gamma ray. Hence,
the entropy decreases as the scale increases. Also it is shown that
the entropy of NPHI is small at small scales which proves the
existence of local correlation, but increases for large scales. This
expresses the separated and uncorrelated porosity. Comparing Fig.$2$
and Fig. $3$ we deduce that small entropy corresponds to
non-Gaussian PDF. note that this result is always true specially
when the scale is small.

\begin{figure}[t]
\centering
\includegraphics[width=9cm,height=7cm,angle=0]{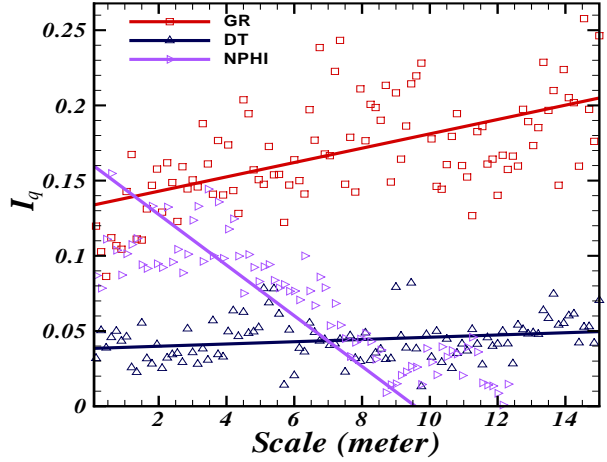}
\caption{Relevance between difference information of petrophysical
quantities contain GR (square), DT (delta), NPHI (right triangle)
and scale according to Eq. $8$. Solid lines display linear fitting
corresponding to the mentioned quantities.} \label{fig4}
\end{figure}

In order to show the multifractal characteristic of the data sets
the Tsalis difference information needs to be calculated \cite{n,o}
\begin{equation}
I_{q} (p:p_{0}) =\frac {1} {1-q} [1-\int{p^{q} p_{0} ^{1-q} dx}],
\end{equation}
where transition is performed between states $p$ and $p_{0}$. Since
the generalized entropy $S_{q}$ (Eq. $(2)$) has been introduced
based on the concept of multifractals, so the degree of
multifractality corresponds to the information evolution. Actually,
because the phase space of a multifractal geometry is occupied
heterogeneously, we expect high information rate in this system. In
Fig.$4$, it is shown that the difference information received from
GR is high and increasing when the scale becomes greater, that is a
hallmark of strong multifractality across the scales. On the other
hand, the difference information for NPHI is rapidly descending,
meaning that the information at large scales is lost. Note that weak
difference information for DT suggests that DT may be nearly
monofractal \cite{s,t}.

\section{Connection between the non-Gaussian degree and intermittency}

To describe the PDF of the velocity difference between two points in
fully developed turbulent flows, Castaing et al. \cite{p} introduced
the following equation based on a log-normal cascade model \cite{q}:

\begin{equation}
P_{\lambda , \sigma_{0}} (x)=\int_{0} ^{\infty} \frac {1}
{\sqrt{2\pi}\lambda} exp(-\frac {\ln^{2} (\sigma/\sigma_{0})}
{2\lambda^{2}})\times \frac{1} {\sqrt{2\pi}\sigma} exp(\frac
{-x^{2}} {2\sigma^{2}}) d(\ln \sigma),
\end{equation}
where $\lambda$, $\sigma_{0}$ are positive parameters. By taking the
limit $\lambda \rightarrow 0$ in Eq. $(12)$, a Gaussian distribution
is obtained. As verified in \cite{r}, the shape of $P_{\lambda}$ is
mainly determined by $\lambda^{2}=\langle (\delta \ln
\sigma)^{2}\rangle$. In fact $\lambda^{2}$ quantifies how fatal the
non-Gaussianity of data sets are \cite{u,w,w1}. As in the case where
$q$ determined the non-Gaussian property, see Eq. $(6)$.

In Fig.$5$ the behavior of the entropy for data sets is plotted vs.
$q$ and $\lambda^{2}$ in the left and right panels, respectively.
since entropy is a measure of lack of information, the system with
low entropy is eager to receive more information potential. This
reveals that the non-Gaussian factor $\lambda^{2}$ is a sign of a
sudden phase transient in the system. Meaning that the system has
two phases which is a logical reason to get close to a petroleum
reservoir. As seen for Gamma ray series in Fig.$5$, at high
$\lambda^{2}$ the entropy tends to decrease to show the existence of
global correlation with a strong interaction with environment. In
fact the increase in the $\lambda^{2}$ leads to a non-Gaussian PDF.
This confirms a high $q$ which results in having valuable
information about the quantity. In contrast, there is high value of
entropy at low $\lambda^{2}$ for NPHI which may be due to the
Gaussian behavior and loss of the information of porosity.

\section{Conclusions}

There are systems in which long-ranged correlations give rise to
strong interactions with environment and violate ergodicity. These
systems can not be described by classical BG statistics. Thus
another definition for entropy which is the basis of statistical
mechanics was defined by Tsallis in $1988$. In this research the
statistical mechanics based on the approach introduced by Tsallis
was applied for analyzing petrophysical quantities which contain
Gamma emission (GR), sonic transient time (DT) and Neutron porosity
(NPHI). Based on the Tsallis entropy, we fitted the data sets to the
probability density function (PDF) in order to obtain the
non-Gaussian degree, $q$ of each series which is able to describe
any system at any scale.

There exist various indicators to analysis the petroleum reservoirs.
In this article, we have investigated the non- Gaussianity (uncertainly)
in some indicators which could be appeared near the interface by two
environments. Then, compared this effect in some indicators in various
scales and measures the information contained in each indicators.

In Fig. $3$ we showed that large degree of $q$ relates to a
fat-tailed PDF in Fig. $2$ which exhibits global correlation in the
formation of the system. For the GR the fat tailed PDF was shown,
however, for NPHI the fat tailed PDF was observed at small scales.
This is due to the fact that the local correlation in NPHI exists.
Hence, for DT and NPHI, $q$ and the efficiency of non-Gaussianity
decreases.

Non-Gaussian PDF at large scales gives us information about the
criticality in the system which results in the entropy decreases.
The extremely increase of entropy for NPHI at large scales is due to
the Gaussian behavior of PDF which causes separated and uncorrelated
porosity. The non-Gaussian factor (intermittency) $\lambda^{2}$,
which is also another important quantity is obtained from the
Castaing equation. High $\lambda^{2}$ is a sign of two phases in the
system which corresponds to a fat-tailed PDF or in other words a
large $q$. This means that high $\lambda^{2}$ for GR at all scales
gives us valuable information about the existence of global
correlation and a sudden phase transition in this series. The
difference information of the data sets was shown in Fig. $4$. we
showed that GR has a high difference information specially at large
scales. In addition, it was shown that GR exhibits a multifractal
property, which is due to the fact that a high information rate in a
system is present. While for DT, a monofractal property was
exhibited which is due to the low information rate.

\begin{figure}[t]
\centering
\includegraphics[width=12.5cm,height=5.5cm,angle=0]{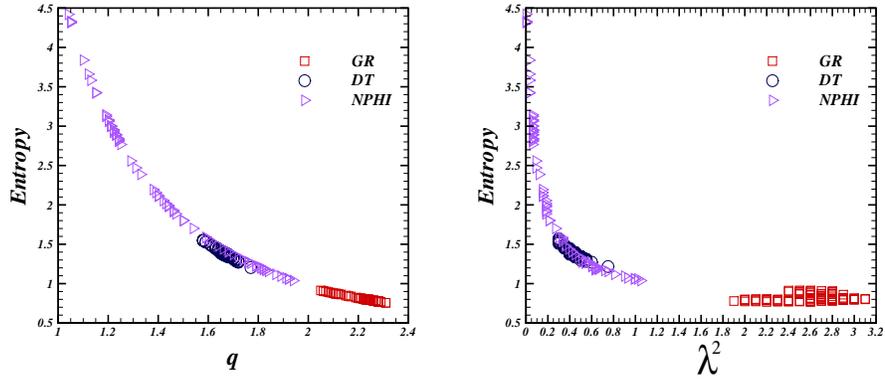}
\caption{illustration of the entropy (Eq. ($2$)) for the data sets:
GR (square), DT (delta), NPHI (right triangle) vs. non-Gaussian
degree, $q$ in the right panel and the intermittency $\lambda^{2}$
defined by Castaing equation in log-normal cascade model (Eq.
($9$))in the left panel.}\label{fig5}
\end{figure}
%@@@@@@@@@@@@@@@@@@@@@@@@@@@@@@@@@@@@@@@@@@@@@@@@@@@@@@@@
%@@@@@@@@@@@@@@@@@@@@@@@@@@@@@@@@@@@@@@@@@@@@@@@@@@@@@@@@

\end{document}